\documentclass[12pt]{article}
\usepackage{amsfonts}
\usepackage{amsmath}
\usepackage{amscd}

\begin{document}
\title{\bf { Spectrum generating algebra for the continuous spectrum of a free particle in Lobachevski space}}

\author{M. Gadella$^1$, J. Negro$^1$, G.P. Pronko$^2$, M. Santander$^1$.}

\maketitle

\noindent $^1$ Departamento de F\'{\i}sica Te\'orica, At\'omica y
\'Optica, Facultad de Ciencias, Universidad de Valladolid, 47005
Valladolid, Spain

\noindent $^2$ Department of Theoretical Physics, IHEP. Protvino,
Moscow Region 142280, Russia.

\begin{abstract}

In this paper, we construct a Spectrum Generating Algebra (SGA) for
a quantum system with purely continuous spectrum: the quantum free
particle in a Lobachevski space with constant negative curvature.
The SGA contains the geometrical symmetry algebra of the system plus
a subalgebra of operators that give the spectrum of the system and
connects the eigenfunctions of the Hamiltonian among themselves. In
our case, the geometrical symmetry algebra is $\frak{so}(3,1)$ and
the SGA is $\frak{so}(4,2)$. We start with a representation of
$\frak{so}(4,2)$ by functions on a realization of the Lobachevski
space given by a two sheeted hyperboloid, where the Lie algebra
commutators are the usual Poisson-Dirac brackets. Then, introduce a
quantized version of the representation in which functions are
replaced by operators on a Hilbert space and Poisson-Dirac brackets
by commutators. Eigenfunctions of the Hamiltonian are given and
``naive'' ladder operators are identified. The previously defined
``naive'' ladder operators shift the eigenvalues by a complex number
so that an alternative approach is necessary. This is obtained by a
non self-adjoint function of a linear combination of the ladder
operators which gives the correct relation among the eigenfunctions
of the Hamiltonian. We give an eigenfunction expansion of functions
over the upper sheet of two sheeted hyperboloid in terms of the
eigenfunctions of the Hamiltonian.

\end{abstract}

\section{Introduction}

The notion of the Spectrum Generating Algebra (SGA) was introduced
many years ago by Barut and Bohm \cite{BB} and independently by
Dothan, Gell-Mann and Neeman \cite{D} for the construction of
multiplets in elementary particle theory. The notion of SGA in
quantum mechanics is suitable for the construction of the Hilbert
space of states for a given system using representation theory. The
point of departure is the geometrical symmetry group for a given
system. The representations of this algebra give the subspace of the
whole Hilbert space of eigenstates corresponding to a fixed energy.
Then, we need to add some generators to the algebra so that the new
elements, the ladder operators, connect states of different
energies. This new generators and hence the ladder operators cannot
commute with the Hamiltonian of the system. Thus,  the SGA will
generate the {\it whole} Hilbert space of eigenfunctions starting
from just one eigenfunction and following some prescriptions on  the
application of the operators of the algebra.

In a former publication \cite{P1}, we have discussed the
construction of the SGA for the free particle in the three
dimensional sphere, $S^3$, where the Hamiltonian has a pure discrete
spectrum. In that case, the initial space isometry algebra or
geometrical algebra was $\frak{so}(4)$, while the SGA that we
constructed was isomorphic to $\frak{so}(4,2)$.

The objective of this paper is to explore the possibility of
extending the notion of SGA for systems with purely continuous
spectrum. A typically non-trivial  example in which this situation
arises is in the three dimensional Lobachevski space. Then, our aim
was constructing a SGA for the free particle on a space of negative
constant curvature, which can be realized as the upper sheet of a
two sheeted hyperboloid ${\cal H}^3$ embedded in the Minkowskian
space ${\mathbb R}^{3+1}$.

In this situation, the geometrical algebra is $\frak{so}(3,1)$ and
we shall show that the SGA is again $\frak{so}(4,2)$. However, the
situation is quite different than in the previous study case
concerning the free motion in $S^3$  where the operator that
parameterizes the Hamiltonian is a generator of a compact subgroup
of $SO(4,2)$, while the analogous operator for ${\cal H}^3$ does not
have this property. In the situation under our study, we do not use
a maximal compact subalgebra in order to construct the basis, but
instead a subalgebra including generators of noncompact subgroups.
Then, ladder operators can be expected to be functions of generators
of the algebra not in $\frak{so}(3,1)$.

Once we have constructed ladder operators for the free particle in
the two-sheeted hyperboloid, a somehow unexpected situation emerges:
the naive choice for ladder operators that should have served to
construct the Hilbert space supporting the representation change the
energy by a complex number. This result means that the ladder
operators take any vector out of the Hilbert space. This illness has
a remedy, which is the construction of a complex power of certain
linear combinations of ladder operators. This action will solve the
problem at the same time that it creates a bridge between quantum
theory in Lobachevski space and the Gelfand-Graev transformation
\cite{GGV}.

We have organized this paper as follows: In Section 2, we construct
the generators of the Lie algebra $\frak{so}(4,2)$ corresponding to
either a free classical particle in a one or two sheeted hyperboloid
and give their relations in terms of Dirac brackets. In Section 3,
we construct the quantized version of the material introduced in
Section 2 including the restrictive relations necessary for the
determination of an irreducible representation of the algebra. We
define the ladder operators and give relations between ladder
operators and other generators of the algebra. Finally, in Section
4, we restrict our study to the three dimensional Lobachevski space
realized by one sheet of the two sheeted hyperboloid ${\cal H}^3$.
Here, we obtain a generalization of plane waves for the free
particle on the hyperboloid. We find that the previously defined
ladder operators shift the energy of these plane waves by a complex
number so that a new concept of ladder operators are defined to
correct this anomaly. The construction of the SGA for the free
particle in ${\cal H}^3$ is then complete. This paper closes in
Section 5, showing an eigenfunction expansion of functions on the
two sheeted hyperboloid in terms of generalized pane waves.

\section{A classical particle in a three dimensional hyperboloid.}

We consider the two sheeted three dimensional hyperboloid ${\mathcal
H}^3$ immersed into an ambient Minkowskian space ${\mathbb R}^{3+1}$
with equation $x^ix^j\,g_{ij}=(x^1)^2+(x^2)^2+(x^3)^2-(x^4)^2=-1$,
 which can be written in
the usual shorthand form as $x^2=x_ix^i=- 1$. Note that here Latin
indices will run from 1 to 4, that the metric is $g_{ij}={\rm
diag}(1,1,1,-1)$ and that we have to sum over repeated indices from
1 to 4. Henceforth, we shall use the standard convention relative to
the operations of lowering and raising indices using the metric
$g_{ij}$.

Now, let us consider the Lagrangian of the free particle with mass
$m=1$ defined in the ambient space ${\mathbb R}^{3+1}$, $L:=\frac12
\,\dot x^i\dot x^j\,g_{ij}$.

Then, its restriction to the hyperboloid ${\cal H}^3$  is given by

\begin{equation}\label{1}
L=-\frac14(\dot x^ix^j-\dot x^jx^i)\, (\dot x_i x_j-\dot x_j x_i)\,,
\end{equation}
where the dot means derivative with respect time. The global minus
sign comes from the condition $x^ix_i=-1$ defining ${\mathcal H}^3$.

The canonical momenta are determined by

\begin{equation}\label{2}
p_i:=\frac{\partial L}{\partial \dot x^i}=-(x^k\dot x^j-x^j\dot
x^k)x_j\,g_{ik}\, ,
\end{equation}
and satisfies the primary constraint
\begin{equation}\label{3}
x^ip_i=x^ip^j\,g_{ij}=0\,.
\end{equation}

The Legendre transformation of the Lagrangian (\ref{1}) gives the
canonical Hamiltonian for the free motion in ${\mathcal H}^3$ as:
\begin{equation}\label{4}
H=-\frac12\;J_{ij}J^{ij}\,,\qquad J^{ij}=x^ip^j-x^jp^i\,,
\end{equation}
where each $J^{ij}$ has the structure of an angular momentum. Our
strategy to work in ${\cal H}^3$ will be the following: Instead of
dealing in the 8-dimensional phase space with dynamical variables
$x^i,p_i$ satisfying the canonical Poisson brackets
$\{x^i,p_j\}=\delta^i_j$, we impose the primary constraint (\ref{3})
and the gauge fixing condition
\begin{equation}\label{5}
x_ix^i=- 1\,.
\end{equation}
According to the usual procedure \cite{DI,guru}, we also introduce
the Dirac brackets

\begin{equation}\label{dirac}
\{x^i,x^j\}_D=0\,,\quad    \{p_i,x^j\}_D=\delta^j_i-x^ix_j\,, \quad
\{p_i,p_j\}_D=J_{ij}\,.
\end{equation}
In the sequel, we prefer to use the variables $x^i$ and $p_i$
subject to the Dirac brackets (\ref{dirac}) instead of defining a
set of independent variables in the hyperboloids. Therefore, from
now on, as we shall work in the configuration space ${\cal H}^3$
where only Dirac brackets will be appropriate, we suppress the label
$D$, such as it appears in (\ref{dirac}).

 Using (\ref{dirac}), we
obtain:
\begin{eqnarray}
\{ J_{ij},J_{kl}\} &=& g_{ij}\,J_{kl}+g_{jk}\,J_{il}-g_{ik}\,J_{jl}-g_{jl}\,J_{ik}\,,    \nonumber \\[2ex]
  \{J_{ij},x_l\}&=& g_{jl}\,x_i-g_{il}\,x_j\,. \label{2.7}
\end{eqnarray}
Therefore, the generators $J_{ij}$ span the geometrical symmetry
group for the hyperboloid, $SO(3,1)$. Its Casimirs are
\begin{equation}\label{2.1}
C=\frac12\,J_{ij}J^{ij}\,,\qquad {\rm and}\qquad \widetilde
C=\epsilon^{ijkl}\,J_{ij}J_{kl}\,.
\end{equation}
With this realization, we have that $\widetilde C=0$ and $C$
coincides up to a sign with the Hamiltonian derived from the above
Lagrangian. The fact that we are moving on the hyperboloid is
characterized by the constraint condition (\ref{3}) plus the gauge
condition (\ref{5}).

Using (\ref{2.7}), we can obtain the following commutators:

\begin{equation}\label{2.8}
\{C,x_i\}=-2J_{ik}x^k\,,\qquad \{C,J_{ik}x^k\}=2Cx_i\,.
\end{equation}
Since $H=-C$ and $H$ is given by formula (\ref{2.5}), then we
conclude that $-C$ is  positive, one may denote by $\sqrt{-C}$ its
unique positive square root. Note that $H=-C$ shows that the
Hamiltonian $H$ is positive.

Then, using the above commutators we obtain the following new ones:
\begin{eqnarray}
 \{\sqrt{-C},x_i\,\sqrt{-C}\} &=& J_{ik}\,x^k\,,  \label{2.9} \\[2ex]
\{ \sqrt{-C},J_{ik}\sqrt{-C}\}  &=& \sqrt{-C}\,x_i\,, \label{2.10}
\\[2ex]
\{ \sqrt{-C}\,x_i,\sqrt{-C}\,x_j\} &=&  \,J_{ij}\,, \label{2.11}
\\[2ex]
\{J_{ik}\,x^k, J_{jl}\,x^l\} &=& -J_{ij}\,,  \label{2.12} \\[2ex]
\{\sqrt{-C}\,x_i, J_{jk}\,x^k\} &=& -\sqrt{-C}\, g_{ij} \,.
\label{2.13}
\end{eqnarray}
Then, if we use the following notation:

\begin{equation}\label{2.14}
 M_{ij}:=J_{ij}\,,\quad M_{5i}:=\sqrt{-C}\;x_i\,,\quad M_{6i}:=
 J_{ik}\,x^k\,,\quad M_{56}:=\sqrt{-C}\,,
\end{equation}
we note that the $\{M_{ab}\}$ satisfy the following commutation
relations (henceforth, indices $a$, $b$ will run out from 1 to 6):

\begin{equation}\label{2.15}
   \{M_{ab},M_{cd}\}=g_{ab}M_{cd}+
  g_{bc}M_{ad}-g_{ac}M_{bd}-g_{bd}M_{ac}\,,
\end{equation}
where the metric $g_{ab}$ is given by

\begin{equation}\label{2.16}
 g_{ab}={\rm diag}\,(1,1,1,-1,-1,1)\,.
\end{equation}
Observe that the metric $g_{ab}$ has the signature $(4,2)$. This
fact and the explicit form of the commutation relations (\ref{2.15})
shows that the $\{M_{ab}\}$ are the generators of the algebra
${\frak so}(4,2)$.

It is also important to remark that in the realization given by
(\ref{2.14}) the following relations hold:

\begin{equation}\label{2.17}
 T_{ab}:=M_{ac}M_{bd}g^{cd}=0\qquad {\rm and} \qquad
 R^{ab}:=\epsilon^{abcdef}\,M_{cd}M_{ef}=0\,,
\end{equation}
where $\epsilon^{abcdef}$ is the completely antisymmetric tensor.
Relations (\ref{2.17}) are called the {\it restrictive relations}
for this representation of the algebra $\frak{so}(4,2)$. These
relations do not change under the action of the algebra,
 since a direct calculation using (\ref{2.15}) shows
that $\{M_{ab}\}$ and $\{R^{cd}\}$ are $\frak{so}(4,2)$ two-tensors.
The situation is in complete analogy to the similar problem on $S^3$
already studied in \cite{P1}.

\section{Quantum SGA}

Next, we are going to introduce the quantum version of the previous
study. If in Section 2, we have given a representation of the
algebra ${\frak so}(4,2)$ suitable for a description of the free
particle on the hyperboloid ${\cal H}^3$, now  we proceed by giving
a representation of this algebra such that its elements are
operators on a Hilbert space. To implement this objective, we
transform Dirac brackets (\ref{2.15}) into commutators, which have
the following form:

\begin{equation}\label{3.1}
[M_{ab},M_{cd}]=-i(g_{ab}M_{cd}+
  g_{bc}M_{ad}-g_{ac}M_{bd}-g_{bd}M_{ac})\,,
\end{equation}
where we require the $M_{ab}$ to be Hermitian operators on a
suitable Hilbert space. In the sequel, we shall use the following
standard notation:
\begin{eqnarray}
J_{ij}:= M_{ij}\,,\qquad i,j=1,\dots,4\,, \nonumber \\[2ex]
K_i:= M_{5i}\,,\qquad  L_i:= M_{6i}\,, \qquad  h:= M_{56}\,.
\label{3.2}
\end{eqnarray}
In matrix form, we can write
\begin{equation}\label{3.3}
M_{ab}=\left(
\begin{array}{ccc}
J_{ij} & K_i & L_i \\[2ex]
-K_i & 0 &  h \\[2ex]
-L_i & - h & 0 \\
         \end{array}
       \right)\,, \qquad i,j=1,\dots,4\,.
\end{equation}
Commutation relations (\ref{3.1}), along with definitions
(\ref{3.2}), give explicitly:
\begin{eqnarray}
[J_{ik},J_{lm}]=-i(g_{im}J_{kl}+g_{kl}J_{im}-g_{il}J_{km}-g_{km}J_{il})\nonumber
\\[2ex]
[J_{ik},K_l]=-i(g_{kl}K_i-g_{il}K_k),\quad
[J_{ik},L_l]=-i(g_{kl}L_i-g_{il}L_k)\nonumber
\\[2ex]
[K_i,K_j]=-[L_i,L_j]=i- J_{ij},\quad [K_i,L_j]=i g_{ij}h\nonumber
\\[2ex]
[h,K_i]=-iL_i,\quad[h,L_i]=-iK_i\,. \label{3.4}
\end{eqnarray}
Note that $g_{55}$ and $g_{66}$ have opposite sign. Then, the
generator $M_{56}$ has always hyperbolic character. In consequence,
$h$ is a noncompact generator corresponding to hyperbolic rotations.
It can be easily shown that

\begin{equation}\label{3.5}
H= -\frac12\,J_{ij}J^{ij}=1+h^2\,.
\end{equation}

At this point it is interesting to note that as is well known
\cite{O}, the spectrum of $C=\frac12\,J_{ij}J^{ij}$ is given by
$-1+\zeta^2$, where $\zeta$ runs either into the real interval
$[-1,1]$ or into the imaginary axis. This shows that the spectrum of
$C$ is non positive and therefore $C\le 0$. Correspondingly,
$SO(3,1)$ has two series of unitary irreducible representations: the
principal series labeled by values of $\zeta=i\rho$ with $\rho\in
(-\infty,\infty)$ and the supplementary series labeled by
$\zeta\in[-1,1]$. The spectrum of $C$ has the form $-1-\rho^2\le -1$
in the first case and $-1<-1+\zeta^2<0$ in the second.

\subsection{Restrictive relations}

We have shown in \cite{P1} the importance of the quantum version of
the restrictive relations in order to fix the representation of the
algebra $\frak{so}(4,2)$. These restrictive relations are the
symmetrized version of (\ref{2.17}) and can be written in the
following form:
\begin{eqnarray}
T_{ab}=(M_{ad}M_{be}+M_{be}M_{ad})g^{de}+cg_{ab} =0\,,\label{3.6}
\\[2ex]
 R^{ab}=\varepsilon^{abcdef}(M_{cd}M_{ef}+M_{ef}M_{cd})=0\,,
 \label{3.7}
\end{eqnarray}
where $c$ is a constant to be determined later. In terms of the
notation proposed in (\ref{3.2}) and (\ref{3.7}), $T_{ab}=0$ is
equivalent to the following set of equations:
\begin{eqnarray}
{T}_{ij}= J_{ik}J_j^k+J_j^kJ_{ik}-(K_i K_j+K_j K_i)+(L_i L_j+L_j
L_i)+c g_{ij}=0\,,
\label{3.8}\\[2ex]
{T}_{5i}= (h L_i+L_i h)-(J_{ij}K^j+K^jJ_{ij})=0\,,\label{3.9}\\[2ex]
{T}_{6i}= h K_i+K_i h-(J_{ij}L^j+L^jJ_{ij})=0\,,\label{3.10}\\[2ex]
{T}_{56}= K_i L^i+L^i K_i=0\,,\label{3.11}\\[2ex]
{T}_{55}= 2(K_i^2+ h^2)- c=0\,,\label{3.12}\\[2ex]
{T}_{66}=2( L_i^2- h^2)+ c=0\,. \label{3.13}
\end{eqnarray}

Analogously, for $R^{ab}=0$, we have
\begin{eqnarray}
R^{ij}=0\Longrightarrow K_i L_j+L_j K_i-(L_i K_j+K_j L_i)-2 h
J_{ij}=0\,,\label{3.14}\\[2ex]
R^{5i}=0\Longrightarrow
\varepsilon^{ijkl}(L_jJ_{kl}+J_{kl}L_j)=0\,,\label{3.15}\\[2ex]
R^{6i}=0\Longrightarrow
\varepsilon^{ijkl}(K_jJ_{kl}+J_{kl}K_j)=0\,,\label{3.16}\\[2ex]
R^{56}=0\Longrightarrow \varepsilon^{ijkl}J_{ij}J_{kl}=0\,.
\label{3.17}
\end{eqnarray}

The space ${\cal H}$ supporting this representation of the algebra
${\frak so}(4,2)$ is given by the vectors $\psi$ such that
$T_{ab}\psi=0$ and $R^{ab}\psi=0$.

The algebra ${\frak so}(4,2)$,  together with the above restrictive
relations, is the SGA for the quantum free motion on ${\cal H}^3$.
In order to justify this terminology, we need to define creation and
annihilation operators (although the spectrum for the free particle
in our case cannot be expected to be discrete). As we did in the
case of the free particle in $S^3$ \cite{P1}, a natural choice seems
to be
\begin{equation}\label{3.19}
   A_i^\pm:=K_i\pm L_i\,,\qquad i=1,2,3,4\,.
\end{equation}

Since $K_i$ and $L_i$ should be Hermitian so they are $A_i^\pm$.
These operators satisfy the following important commutation
relations:
\begin{eqnarray}
A_i^+(A^i)^+ &=& A_i^-(A^i)^-=0\,,\quad i=1,2,3,4\,, \label{3.20}
\\[2ex]
[A_i^+,A_j^+] &=& [A_i^-,A_j^-]=0 \,,\quad i,j=1,2,3,4\,,
\label{3.21}
\\[2ex]
[A_i^+,A_j^-] &=& -2i (J_{ij}+g_{ij}h)\,, \quad
i,j=1,2,3,4\,.\label{3.22}
\end{eqnarray}
We write separately the next commutation relations due to their
 importance:
\begin{equation}\label{3.23}
hA_j^+=A_j^+(h-i)\,,\qquad hA_j^-=A_j^-(h+i)\,,\qquad j=1,2,3,4\,,
\end{equation}
where $i$ is here the imaginary unit.

In principle, the operators $A_i^\pm$ are the naive equivalent of
the ladder operators defined for $S^3$ in \cite{P1}. However, this
procedure by analogy does not work here. This is due to the
appearance of the term $i$ in (\ref{3.23}) as we shall see later.
Note that $A_i^\pm$ as well as their real linear combinations belong
to the algebra of generators of $SO(4,2)$, so that for any unitary
irreducible representation of $SO(4,2)$, real linear combinations of
$A_i^\pm$ are represented by means of self adjoint operators on the
Hilbert space supporting this representation.

Now, we look for quantum operators corresponding to the classical
coordinates $x^i$. These operators can now be defined as

\begin{equation}\label{3.24}
  X_i:=f(h)(A_i^++A_i^-)f(h)\,,\qquad i=1,2,3,4\,,
\end{equation}
where $f(h)$ is a function of $h$ that we determine by the
hypothesis that $[X_i,X_j]=0$ and that $X_iX^i$ is a c-number (which
we shall choose to be $\pm1$). These two conditions not only
determine $f(h)$, but also the number $c$ in (\ref{3.8}) that
happens to be $c=2$. The function $f(h)$ is

\begin{equation}\label{3.25}
 f(h)=\frac1{\sqrt 2}\,\frac1{\sqrt h}\,,
\end{equation}
which after (\ref{3.19}) and (\ref{3.24}) implies that

\begin{equation}\label{3.26}
  X_i=\frac1{\sqrt h}\,K_i\,\frac1{\sqrt h}\Longleftrightarrow K_i=\sqrt h\,X_i\,\sqrt
  h\,,\quad
  i=1,2,3,4\,.
\end{equation}
The conditions we have enforced on $X_i$ correspond to the
requirement that they are coordinates of the ambient Lobachevski
space $x^ix_i=-1$ and also that they satisfy the proper gauge
condition on the hyperboloid.

Now, let us find an interesting relation for $L_i$ in terms of
$X_i$, $J_{ik}$ and $h$. From (\ref{3.19}), it readily follows the
first identity in (\ref{3.23}) that

\begin{equation}\label{3.27}
hL_i+L_ih=\frac12\,\{h(A_i^+-A_i^-)+(A_i^+-A_i^-)h\}=:\frac12\,g(h)(A_i^+-A_i^-)g(h)\,,
\end{equation}
while the second comes from the commutation relations (\ref{3.4})
involving $h$. The function $g(h)$ should be determined from those
relations. Equation $g(h)$ should fulfil the equation

\begin{equation}\label{3.28}
g(h)g(h+i)=2h+i\,.
\end{equation}
Equation (\ref{3.28}) can be solved after some work and gives as
solution:

\begin{equation}\label{3.29}
g(h)=2\left[  \frac{\Gamma\left( \frac{ih}2+\frac 34
\right)}{\Gamma\left( \frac{ih}{2}+\frac 14
\right)}\;\frac{\Gamma\left( -\frac{ih}2+\frac 34
\right)}{\Gamma\left( -\frac{ih}{2}+\frac 14 \right)} \right]^{1/2}
\end{equation}
Due to the fact that $\Gamma(z^*)=\Gamma^*(z)$, where the star
denotes complex conjugation, the function $g(h)$ is always positive
(we take the principal branch in the square root).

Now, combining (\ref{3.27}), (\ref{3.19}) and the restrictive
relation (\ref{3.9}), we find that

\begin{equation}\label{3.30}
L_i=g^{-1}(h)\,\sqrt h\,(J_{ik}\,X^k+X^kJ_{ik})\,\sqrt
h\,g^{-1}(h)\,.
\end{equation}

At this point, we have completed the first task: we have constructed
all the generators of the SGA as functions of the operators $h$,
$X_i$ and the geometrical generators $J_{ij}$. In particular $L_i$
and $K_i$ are given by formulas (\ref{3.26}) and (\ref{3.30}). In
the next section, we shall construct the space of states of our
system with the help of the SGA.

\section{The quantum free Hamiltonian on ${\cal H}^3$}

Hereafter, we shall use the following notation:
$x_\alpha^2=x_1^2+x_2^2+x_3^2$, and  $\dot x_\alpha x^\alpha=\dot
x_1x_1+\dot x_2x_2+\dot x_3x_3$. Note that Greek indices $\alpha$
and $\beta$ run from 1 to 3 and that for these indices the
distinction between upper and lower makes no sense. From the
hyperboloid equations, $x_ix^i=- 1$, we obtain that
$x_4=\sqrt{x_\alpha^2+ 1}$. Using this chart of coordinates for
${\cal H}^3_1$, we obtain the following new expression for the
classical Lagrangian (\ref{1}),
\begin{equation}\label{2.4}
L=\frac12\;\left(\dot x_\alpha^2-\frac{(\dot x_\alpha x_\alpha
)^2}{x_\alpha^2+ 1}\right)\,.
\end{equation}
This Lagrangian is written in terms of the coordinates $x_\alpha$.
Its Legendre transformation (with respect to the independent
variables $x_\alpha$, $\alpha=1,2,3$) gives the Hamiltonian:
\begin{equation}\label{2.5}
H= \frac12\; (p_\alpha^2+(x_\alpha p_\alpha)^2)\,,
\end{equation}
where $x_\alpha$ and $p_\beta$ are the classical canonical conjugate
coordinates of the position and momentum respectively.

Then, let us proceed with the quantization of this system.

For ${\cal H}^3$, the Hilbert space of states will be  the space of
all Lebesgue measurable functions on the hyperboloid with a metric
which is the restriction of the Lebesgue measure on ${\mathbb
R}^{3+1}$ to the hyperboloid. This is

\begin{equation}\label{3.31}
 \langle \varphi|\psi\rangle:=\int d^4
 x\,\delta(x_ix^i+1)\,\varphi^*(x)\,\psi(x)=\int_{{\mathbb R}^3}
 \frac{d^3x_\alpha}{\sqrt{{\bf x}^2+1}}\,\varphi({\bf x})\,\psi({\bf
 x})\,,
\end{equation}
where ${\bf x}=(x_1,x_2,x_3)$, $\alpha=1,2,3$\,.

Canonical quantization of the classical free Hamiltonian (\ref{2.5})
gives us a quantum Hamiltonian having the same expression as
(\ref{2.5}), by replacing $ x_\alpha$ and $p_\beta$ by a pair of
canonical conjugate operators for the components of the position
$X_\alpha$ and momentum $P_\beta$ respectively. Note that $P_\alpha$
is not just partial derivation with respect to to $x_\alpha$
multiplied by $i$, since this operator is not Hermitian with respect
to the scalar product (\ref{3.31}). Instead, we should define
\begin{equation}\label{3.32}
  P_\alpha:= ({\bf X}^2+1)^{1/4}\;(-i\partial_\alpha)\; ({\bf
X}^2+1)^{-1/4}\,,
\end{equation}
where $\partial_\alpha$ is the partial derivative with respect to
$x_\alpha$ and ${\bf X}=(X_1,X_2,X_3)$. These $P_\alpha$ together
with $X_\alpha$ (defined as multiplication operators) satisfy the
canonical commutation relations
$[X_\alpha,P_\beta]=i\delta_{\alpha\beta}$. Further, we have to
remark that the operators $X_i$, $i=1,2,3,4$ are realizations of the
coordinate operators in the ambient space defined in the previous
subsection since they satisfy the same commutation relations and the
same gauge condition $x_ix^i=-1$. Then, we shall use capital letters
to denote these operators $X_i$ in the sequel. From this expression
for the $P_\alpha$, we can obtain the Hermitian version of the
classical Hamiltonian (\ref{2.5}) as

\begin{equation}\label{3.33}
H=-\frac 12\, \sqrt{{\bf
X}^2+1}\;\partial_\alpha\;\frac{\delta_{\alpha\beta}+
 X_\alpha X_\beta}{\sqrt{{\bf X}^2+1}}\;\partial_\beta\,,
\end{equation}
where we sum over the repeated indices $\alpha$ and $\beta$ running
from 1 to 3.

Our next goal is to solve the Schr\"odinger equation

\begin{equation}\label{nueve}
H\psi=\lambda \psi
\end{equation}
associated to this Hamiltonian. By the form of $H$ in (\ref{3.33}),
we see that $H$ cannot have bound states and that the solutions of
$H\psi=\lambda \psi$ are not expected to be normalizable. As the
ambient space is ${\mathbb R}^{3+1}$, we expect these wave functions
to depend on the four dimensional vector $x^i$. We are looking for
those special solutions of the Schr\"odinger equation (\ref{nueve})
which depend on the four variables $x_i$ through the single
combination $f:=k_ix^i=k_\alpha x^\alpha-\sqrt{{\bf x}^2+1}\,k_4$,
$x_4=\sqrt{{\bf x}^2+1}$ showing that the $x_i$ coordinates denote
points in the hyperboloid. In terms of this variable $f$,
(\ref{nueve}) has the form:

\begin{equation}\label{3.34}
 [f^2+({\bf k}^2-k_4^2)]\,\psi''(f)+3f\,\psi'(f)=-\lambda\,\psi(f)\,,
\end{equation}
where the primes indicate derivative with respect to $f$ and ${\bf
k}^2=k_\alpha k^\alpha$. The simplest situation in (\ref{3.34})
happens when ${\bf k}^2-k_4^2=0$. In this case, the general solution
for (\ref{3.34}) is given by

\begin{equation}\label{3.35}
\psi(f)=C_1\,f^{-1+\sqrt{1-\lambda}}+C_2\,f^{-1-\sqrt{1-\lambda}}\,,
\end{equation}
where $C_i$, $i=1,2$ are arbitrary constants. In the general case,
$k_i$ lie on a hyperboloid of the form ${\bf k}^2-k_4^2=-m^2$ ($m^2$
may be positive or negative). Now, the general solution has the form

\begin{equation}\label{3.36}
\psi(f)=C_1\;\frac{(f+\sqrt{f^2-m^2})^{i\rho}}{\sqrt{f^2-m^2}}+
C_2\;\frac{(f+\sqrt{f^2-m^2})^{-i\rho}}{\sqrt{f^2-m^2}}\,,
\end{equation}
where $\lambda=(1+\rho^2)$. Note that in the limiting case $m=0$,
this solution is equal to (\ref{3.35}) that in terms of $\rho$ is

\begin{equation}\label{3.37}
\psi(f)=C_1\,f^{-1+i\rho}+C_2\,f^{-1-i\rho}\,.
\end{equation}
We shall use this notation in the sequel. It is important to insist
that the parameter which appears in the Schr\"odinger equation is
$\lambda$. For each fixed value of $\lambda$, there exists two
linearly independent solutions as shown in (\ref{3.37}). The
relation between $\lambda$ and $\rho$ is given by
$\lambda=(1+\rho^2)$ as given before. On the other hand, we may look
at $\rho$ as the basic parameter. Then, there is a unique solution
for each $\rho\in(-\infty,\infty)$ given by (with $k_i$ in the cone
${\bf k}^2-k_4^2=0$):

\begin{equation}\label{3.38}
 \psi(f)=\psi(x_\alpha,k_\alpha,\rho)=(x_\alpha k_\alpha-\sqrt{{\bf x}^2+1}\,k_4)^{-1+i\rho}\,.
\end{equation}
Then, we have obtained solutions of the Schr\"odinger equation with
Hamiltonian (\ref{3.33}) labeled by $k_\alpha$ and $\rho$, which
play the same role than the components of the momentum for the
standard Euclidean plane waves $\phi(x,p)=e^{i{\bf x}\cdot{\bf p}}$.

Next, let us consider the quantized version of the components of the
antisymmetric tensor $J_{ij}$, $i,j=1,2,3,4$. We have

\begin{equation}\label{3.39}
J_{\alpha\beta}=X_\alpha P_\beta-X_\beta P_\alpha\,,  \qquad
J_{4\alpha}=\sqrt{{\bf X}^2+1}\;P_\alpha\,, \qquad
\alpha,\beta=1,2,3\,.
\end{equation}
The components $J_{ij}$ satisfy the commutation relations for the
generators of $SO(3,1)$. We also can readily show that the relation
between the quantum analog of Hamiltonian (\ref{2.5}) and $J_{ij}$
is given by

\begin{equation}\label{3.40}
H=-\frac 12\,J_{ij}J^{ij}=1+h^2\,,\qquad {\rm with}\qquad
h=M_{56}\,.
\end{equation}

Then, it is clear after (\ref{nueve}) and (\ref{3.40}) that
$h\psi(f)=\rho\psi(f)$. In order to obtain the action of $L_i$ and
$K_i$, we shall use an operator with four components $T_i$,
$i=1,2,3,4$.  If $X_4:=\sqrt{{\bf X}^2+1}$, let us define  $T_i$ by
\begin{equation}\label{3.41}
T_i:=J_{ij}\,X^j+X^j\,J_{ij}\,,
\end{equation}
then,
\begin{equation}\label{3.42}
T_\alpha=i(X_\alpha(2X_\beta\,\partial_\beta+3)+2\partial_\alpha)\,,\qquad
T_4=i\sqrt{{\bf X}^2+1}\;(2X_\beta\partial_\beta+3)\,,
\end{equation}
where $\alpha,\beta=1,2,3$ and we sum over repeated indices. These
four operators are Hermitian with respect to the scalar product
(\ref{3.31}) and so is the operator $T_ik^i$, where we define the
components $k^i$ as above. This is
\begin{equation}\label{3.43}
T_ik^i=i(X_i k^i
(2X_\beta\,\partial_\beta+3)+2k_\alpha\partial_\alpha)\,.
\end{equation}

The action (\ref{3.43}) on the wave functions (\ref{3.38}) is
\begin{eqnarray}
[T_ik^i]\, \psi(x_\alpha,k_\alpha,\rho)= [T_ik^i]\, (x_\alpha
k_\alpha-\sqrt{{\bf x}^2+1}\,k_4)^{-1+i\rho}
[T_ik^i]\,(x_ik^i)^{-1+i\rho}   \nonumber \\[2ex]
= -(2\rho-i)(x_ik^i)^{i\rho}=-(2\rho-i)(x_ik^i)^{-1+i(\rho-i)}=
-(2\rho-i) \psi(x_\alpha,k_\alpha,\rho-i) \,.
\nonumber\\[2ex]  \label{3.44}
\end{eqnarray}
As we can see from (\ref{3.44}), the action of the operator $T_ik^i$
on $\psi(x_\alpha,k_\alpha,\rho)$ shifts $\rho$ by $-i$.

The action of other operators on the wave functions
$\psi(x_\alpha,k_\alpha,\rho)$ can also be given. For instance, take
expression (\ref{3.30}) for the operators $L_i$, which now can be
defined with the new $J_{ij}$ and have the same commutation
relations than those in the previous section (and therefore they
should be identified). We obtain:

\begin{eqnarray}
[L_ik^i]\,
\psi(x_\alpha,k_\alpha,\rho)&=&[L_ik^i]\,(x_ik^i)^{-1+i\rho}  \nonumber\\[2ex]
&=&-\sqrt{\rho(\rho-i)}(g(\rho)g(\rho-i))^{-1}\,[t_i k^i]\,(x_ik^i)^{-1+i\rho}   \nonumber\\[2ex]
  &=&-\sqrt{\rho(\rho-i)}(g(\rho)g(\rho-i))^{-1}(2\rho-i)(x_ik^i)^{i\rho}\,,   \label{3.45}
\end{eqnarray}
where $g(\rho)$ has been given in (\ref{3.29}) and therefore
satisfies relation (\ref{3.28}). Then, if we apply (\ref{3.28}) into
(\ref{3.45}), we finally get:

\begin{eqnarray}
[L_ik^i]\psi(x_\alpha,k_\alpha,\rho)=\,[L_ik^i]\,(x_ik^i)^{-1+i\rho}
\nonumber
\\[2ex]
=-\sqrt{\rho(\rho-i)}(x_ik^i)^{i\rho}
=-\sqrt{\rho(\rho-i)}\psi(x_\alpha,k_\alpha,\rho-i) \,. \label{3.46}
\end{eqnarray}

Following similar procedures, we can obtain the action of the
operator $K_ik^i$ with $K_i$ as in (\ref{3.26}) into
$\psi(x_\alpha,k_\alpha,\rho)$. This gives:

\begin{equation}\label{3.47}
   [K_ik^i]\,\psi(x_\alpha,k_\alpha,\rho)=\sqrt{\rho(\rho-i)}\,\psi(x_\alpha,k_\alpha,\rho-i)\,.
\end{equation}

From relations (\ref{3.19}), we can determine the action of $A_i^\pm
k^i$ into $\psi(x_\alpha,k_\alpha,\rho)$. This is

\begin{eqnarray}
A^-_ik^i\,\psi(x_\alpha,k_\alpha,\rho)=(K_i+L_i)k^i\,
\psi(x_\alpha,k_\alpha,\rho)=0 \,, \label{3.48}
\\[2ex]
  A^+_ik^i\,\psi(x_\alpha,k_\alpha,\rho)
  =(K_i-L_i)k^i\,\psi(x_\alpha,k_\alpha,\rho)=2\sqrt{\rho(\rho-i)}\, \psi(x_\alpha,k_\alpha,\rho-i)
  \,.\label{3.49}
\end{eqnarray}
We discuss the consequences of (\ref{3.49}) in the next subsection.

\subsection{The ladder operators}

Let us go back to equation (\ref{3.23}). Assume that $\psi$ is an
eigenvector of $h$, $h\psi=\rho\psi$. Then, (\ref{3.23}) gives
$h(A_i^+\psi)=A_i^+(h-i)\psi=(\rho-i)(A_i^+\psi)$. Since $A_i^+$ is
Hermitian (and so is $A_i^+k^ i$), this shows that $A_i^+\psi$ (and
also $A_i^+k^ i\psi$) is not in the domain of $h$. Same happens with
$A_i^-\psi$ and $A_i^-k^ i\psi$. Therefore, it is not possible in
principle to use $A_i^\pm$ as the ladder operators for the
eigenvectors of $h$. This is related to the fact that the spectrum
of $h$ is continuous.

Fortunately, this is not the end of the story. In order to find a
clue on how to proceed, let us analyze the simplest case of
$SO(2,1)$. Its Lie algebra ${\frak so}(2,1)$ has generators $l_0$,
$l_1$ and $l_2$ with commutation relations

\begin{equation}\label{3.50}
 [l_1,l_2]=-il_0\,,\qquad [l_0,l_1]=il_2\,,\qquad [l_0,l_2]=-il_1\,.
\end{equation}

The algebra ${\frak so}(2,1)$ has a generator of a compact subgroup
which is $l_0$ (sometimes, generators of compact subgroups are
denoted as compact generators, which does not mean that they are
compact in the ordinary sense of compact operators on Banach
spaces). In order to construct a unitary irreducible representation
of $SO(2,1)$, we may use the subspace spanned by the eigenvectors of
$l_0$. We may write

\begin{equation}\label{3.51}
 l_0|m\rangle=m|m\rangle\,,
\end{equation}
where we have labeled as $m$ the eigenvalues of $l_0$ for
convenience. Ladder operators can be defined in this case as
$A^\pm=l_1\pm il_2$, so that

\begin{equation}\label{3.52}
 A^\pm|m\rangle=a^\pm(m)|m\pm 1\rangle\,,
\end{equation}
where the coefficients $a^\pm(m)$ should be determined by using
commutation relations (\ref{3.50}) and unitarity conditions for the
elements of $SO(3,1)$.

This is a standard procedure, but it would be interesting to
investigate what would happen if instead of $l_0$ we had insisted in
building the same construction with a noncompact operator, say
$l_1$. If the real number $\lambda$ is in the continuous spectrum of
$l_1$ with generalized eigenvector $|\lambda\rangle$ \cite{GG}, we
have

\begin{equation}\label{3.53}
   l_1|\lambda\rangle=\lambda |\lambda\rangle\,.
\end{equation}

Commutation relations (\ref{3.50}) trivially give

\begin{equation}\label{3.54}
  [l_1,l_0\pm l_2]=\mp i(l_0\pm l_2)\,,
\end{equation}
which implies that the vector given by

\begin{equation}\label{3.55}
 |\widetilde \lambda\rangle:= (l_0+l_2)|\lambda\rangle
\end{equation}
should be an (generalized) eigenvector of $l_1$ with eigenvalue
$\lambda-i$:

\begin{equation}\label{3.56}
l_1 |\widetilde \lambda\rangle=(\lambda-i)|\widetilde
\lambda\rangle\,.
\end{equation}
This shows that the subspace spanned by the generalized
eigenvectors\footnote{Roughly speaking, the vectors $\psi$ of this
space should admit a span in terms of the $|\lambda\rangle$ in the
form $\psi=\int
\langle\lambda|\psi\rangle|\lambda\rangle\,d\mu(\lambda)$. Details
in \cite{GG}.} $|\lambda\rangle$ of $l_1$, with $\lambda$ in the
spectrum of $l_1$, cannot support a unitary representation for
$SO(2,1)$. This problem was already discussed in the literature
\cite{guru1}. We shall introduce here another point of view.

This new point of view is the {\it essential point} in the
construction of a SGA for a Hamiltonian with continuous spectrum and
we introduce it as follows:

According to (\ref{3.19}) operators $A_i^\pm$ are Hermitian and so
are $A_i^\pm k^i$. Therefore, a measurable function of a self
adjoint version of $A_i^\pm$ is well defined\footnote{If we use a
unitary irreducible representation of $SO(4,2)$, the elements of the
Lie algebra $\frak{so}(4,2)$ are represented by self adjoint
operators \cite{K} and therefore $A_i^\pm k^i$ can be represented by
self adjoint operators so that $(A_i^\pm k^i)^{-iu}$ can be well
defined via spectral theory \cite{AJS} with a proper choice of a
branch for the logarithm.} according to the spectral representation
theorem. Our goal is to find a workable expression for $(A_i^\pm
k^i)^{-iu}$ with $u$ real and to show that

\begin{equation}\label{3.57}
h(A_i^+k^i)^{-iu}= (A_i^+k^i)^{-iu}(h-u)\,.
\end{equation}
In order to prove this formula, let us note that as a
straightforward consequence of (\ref{3.23}) is that

\begin{equation}\label{3.58}
   h(A_i^+ k^i)^n= (A_i^+ k^i)^n (h-in)\,.
\end{equation}
Then, for any complex number $z$ one has

\begin{equation}\label{3.59}
  h\exp\{-z (A_i^+ k^i)\}= \exp\{-z (A_i^+ k^i)\} (h+iz(A_i^+
  k^i))\,,
\end{equation}
whenever the exponential be correctly defined. As a matter of fact,
the exponential in (\ref{3.59}) is not always defined and it should
be considered as an abbreviate form of writing the formal series

\begin{equation}\label{3.60}
 \sum_{\ell=1}^\infty \frac{(A_i^+ k^i)^\ell}{\ell!}\,.
\end{equation}
In order to prove (\ref{3.57}), we shall make use of the following
integral representation \cite{GR}:

\begin{equation}\label{3.61}
 \int_{0}^{\infty}dz \,\frac{z^{-1+iu}}{\Gamma(iu)}
 e^{-z\mu}=\mu^{-iu}\,,
\end{equation}
where $u$ is a real number. Now replace $\mu$ by $A_i^+ k^i$ in the
above expression and multiply the resulting formal expression by $h$
to the left. After (\ref{3.61}) this gives:

\begin{eqnarray}
h\int_{0}^{\infty}dz\, \frac{z^{-1+iu}}{\Gamma(iu)}e^{-z\mu}=
\int_{0}^{\infty}dz \,\frac{z^{-1+iu}}{\Gamma(iu)}e^{-z\mu}(h+i\mu
z)\nonumber\\[2ex]
=\int_{0}^{\infty}dz\,
\frac{z^{-1+iu}}{\Gamma(iu)}e^{-z\mu}h+\int_{0}^{\infty}dz\,
\frac{z^{-1+iu}}{\Gamma(iu)}e^{-z\mu}i\mu z\,. \label{3.62}
\end{eqnarray}
Again, we use (\ref{3.61}) in the last term of (\ref{3.62}):

\begin{equation}\label{3.63}
\int_{0}^{\infty}dz \, \frac{z^{-1+iu}}{\Gamma(iu)}e^{-z\mu}i\mu
z=i\mu\mu^{-1-iu}\;\frac{\Gamma(iu+1)}{\Gamma(iu)}=-u\mu^{-iu}\,.
\end{equation}
From there, we readily obtain the commutation relations
(\ref{3.57}). If we recall that
$h\,\psi(x_\alpha,k_\alpha,\rho)=\rho\,\psi(x_\alpha,k_\alpha,\rho)$,
it becomes obvious that

\begin{equation}\label{3.64}
 h\, [ (A_i^+k^i)^{-iu}\,
\psi(x_\alpha,k_\alpha,\rho)]=(\rho-u)\,(A_i^+k^i)^{-iu}\,\psi(x_\alpha,k_\alpha,\rho)\,,
\end{equation}
so that $(A_i^+k^i)^{-iu}\,\psi(x_\alpha,k_\alpha,\rho)$ is an
eigenfunction of $h$ with eigenvalue $\rho-u$. Therefore, it must
exist a constant depending on $u$ and $\rho$, $g(u,\rho)$ such that

\begin{equation}\label{3.65}
[ (A_i^+k^i)^{-iu}\,
\psi(x_\alpha,k_\alpha,\rho)]=g(u,\rho)\,\psi(x_\alpha,k_\alpha,\rho-u)\,.
\end{equation}

The function $g(u,\rho)$ satisfies the following properties

\begin{equation}\label{3.66}
 g(0,\rho)=0\,,\qquad g(i,\rho)=2\sqrt{\rho(\rho-i)}\,,
\end{equation}
where the first relation in (\ref{3.66}) is obvious and the second
come from (\ref{3.49}). These relations will be useful in order to
find the final expression for $g(u,\rho)$. Another property for
$g(u,\rho)$ can be obtained from

\begin{eqnarray}
(A^+_i k^i)^{-iv-iu}\, \psi(x_i,k_i,\rho)= g(u+v)\, \psi(x_i,k_i,\rho-u-v)\,, \nonumber \\[2ex]
 (A^+_i k^i)^{-iv-iu}\, \psi(x_i,k_i,\rho) =   (A^+_i k^i)^{-iv}\, (A^+_i k^i)^{-iu}\,
  \psi(x_i,k_i,\rho) \nonumber\\[2ex]
  =g(u,\rho)\,(A^+_i k^i)^{-iv}\, \psi(x_i,k_i,\rho-u)= g(u,\rho)\,
  g(v,\rho-u)\, \psi(x_i,k_i,\rho-u-v)\,. \label{3.67}
\end{eqnarray}
These identities show the following functional identity for
$g(u,\rho)$:

\begin{equation}\label{3.68}
g(u,\rho)\,g(v,\rho-u)=g(u+v,\rho)\,,
\end{equation}
which has the following solution:

\begin{equation}\label{3.69}
 g(u,\rho)=\left[2^{-iu}i^{-2iu}\frac{\Gamma(-i\rho)\Gamma(1-i\rho)}
{\Gamma(-i(\rho-u))\Gamma(1-i(\rho-u))}\right]^{\frac{1}{2}}\,.
\end{equation}

This completes the discussion on the construction of the ladder
operators and their action on the hyperboloid plane waves
$\psi(x_\alpha,k_\alpha,\rho)$. In the next section, we shall
discuss an interesting formula giving an eigenfunction expansion of
functions over ${\cal H}^3_1$.

\section{General properties of the eigenfunctions of the
Hamiltonian}

Here, we start with an infinitely differentiable function $f(x_i)$
on the hyperboloid ${\cal H}^3_1$, with equation $x_ix^i=-1$. The
function $f(x_i)$ can be transformed into the function $h(k_i)$ on
the cone $k_ik^i=0$ by means of the following integral:

\begin{equation}\label{3.70}
  h(k^i)=\int Dx\, f(x_i)\,\delta(x_ik^i-1)\,,
\end{equation}
where $Dx$ represents here the invariant measure on the hyperboloid
(or equivalently the restriction of the Lebesgue measure on the
hyperboloid):

\begin{equation}\label{3.71}
   Dx=\frac{d^3x_\alpha}{x_4}=\frac{d^3x_\alpha}{\sqrt{x^2_\alpha+1}}\,,\quad
   \alpha=1,2,3\,.
\end{equation}

The integral (\ref{3.70}) for functions $f(x_i)$ on the hyperboloid
gives a function on the cone. This type of transformation has been
considered by Gelfand and Graev \cite{GGV} and holds their name
(Gelfand-Graev transformation). This Gelfand-Graev transform has an
inverse which is given by

\begin{equation}\label{3.72}
  f(x_i)=-\frac1{8\pi^2}\int Dk\,h(k^i)\,\delta''(x_ik^i-1)\,,
\end{equation}
where $Dk$ is the measure on the cone given by

\begin{equation}\label{3.73}
 Dk=\frac{d^3k_\alpha}{k^4}\,.
\end{equation}

Then, let us consider the Mellin transform of the function $h(k^i)$,
which is defined as:

\begin{equation}\label{3.74}
 \phi(k^i,\rho)=\int_0^\infty dt\,h(tk^i)\,t^{-i\rho}\,.
\end{equation}
This Mellin transform has the following inversion formula:

\begin{equation}\label{3.75}
   h(k^i)=\frac1{2\pi}\,\int_{-\infty}^\infty
   d\rho\,\phi(k^i,\rho)\,.
\end{equation}

If we replace in (\ref{3.74}) the expression (\ref{3.70}) for
$h(k^i)$, we find

\begin{eqnarray}
\phi(k^i,\rho)=\int_{0}^{\infty} dt \int Dx\, f(x_i)\, \delta(x_i
k^it-1)\,t^{-i\rho}\nonumber
\\[2ex]
 = \int Dx\, f(x_i)\int_{0}^{\infty} dt\, \delta(x_i k^it-1)\,t^{-i\rho}=\int Dx f(x_i)(x_i
  k^i)^{-1+i\rho}\,. \label{3.76}
\end{eqnarray}

Equation (\ref{3.76}) shows that $\phi(k^i,\nu)$ is an homogenous
function on the cone of degree $-1+i\rho$ and therefore it can be
defined by its values on any contour which crosses all generatrices
of the cone. Equation (\ref{3.76}) can be looked as a generalization
of the Fourier transform of the function $f(x_i)$ with respect to
the integral kernel $\{(x_i k^i)^{-1+i\rho}\}$. In fact this
integral kernel is formed up to plane waves in the same way that the
standard Fourier transform has as integral kernel the Euclidean
plane waves $e^{i({\bf x}\cdot{\bf p})}$. Then, we can find the
inverse transformation of (\ref{3.76}) by entering (\ref{3.75}) into
(\ref{3.72}). The result is given under the form of the following
integral:

\begin{equation}\label{3.77}
f(x_i)= -\frac{1}{16\pi^3}\int Dk\int_{-\infty}^{\infty}d\rho\,
\phi(k^i,\rho)\,\delta''(x_i k^i-1)\,.
\end{equation}
We can rewrite the right hand side of (\ref{3.77}) in the following
form:

\begin{equation}\label{3.78}
  f(x_i)=-\frac{1}{16\pi^3}\int_{\infty}^{\infty}dt\int Dk\,
\int_{-\infty}^{\infty}d\rho\, \phi(k^i,\rho)\,\delta''(x_i
k^i-t)\,\delta(t-1)\,.
\end{equation}
After integrating by parts twice with respect to the variable $t$,
we obtain:

\begin{equation}\label{3.79}
 f(x_i)=-\frac{1}{16\pi^3}\int_{\infty}^{\infty}dt\,
 \delta''(t-1)\int Dk\, \int_{-\infty}^{\infty}d\rho\, \phi(k^i,\rho)\,\delta(x_i k^i-t)\,.
\end{equation}
Now, let us make the change of variables given by $k^i\longmapsto
t\widetilde {k^i}$ in the integrand of (\ref{3.79}). The result is

\begin{eqnarray}
\int Dk\int_{-\infty}^{\infty}d\rho\, \phi(k^i,\rho)\,\delta(x_i
k^i-t) =\int D\widetilde{k}\;\; t^2\int_{-\infty}^{\infty}d\rho\,
\phi(\widetilde {k}_it,\rho)\,\delta(x_i \widetilde{k}^it-t)
\nonumber
\\[2ex]
  = \int D\widetilde{k}\;\; t^2\int_{-\infty}^{\infty}d\rho\, \phi(\widetilde
  {k^i},\rho)\,t^{-1+i\rho}\,
\delta(x_i \widetilde{k^i}-1)t^{-1}\nonumber\\[2ex]
= \int D\widetilde{k} \int_{-\infty}^{\infty}d\rho\, \phi(\widetilde
{k^i},\rho)\,t^{i\rho}\,\delta(x_i \widetilde{k^i}-1)\,.
\label{3.80}
\end{eqnarray}
Then, if we use (\ref{3.80}) into (\ref{3.79}) and integrate over
$t$, we obtain the desired representation of $f(x_i)$ in terms of
its generalized Fourier components $\phi(k^i,\rho)$:

\begin{equation}\label{3.81}
 f(x_i)= \frac{1}{16\pi^3}\int_{-\infty}^{\infty}d\rho\, \rho^2\int Dk\,\phi(k^ii,\rho)\,\delta(x_i
 k^i-1)\,.
\end{equation}

Thus, we have found an analogue the Fourier transform for functions
over an hyperboloid. We may wonder on whether it is also an analogue
of the Plancherel formula in this case. Let us consider the
following parametrization for the four vector with components $k^i$:

\begin{equation}\label{3.82}
  k^i=\omega N^i=({\bf n},1)\omega\,, \qquad {\bf n}^2=1\,,
\end{equation}
where $\omega=k^4$. We use $\omega$ in order to get rid of the
index. In this parametrization, the measure $Dk$ has the following
form:

\begin{equation}\label{3.83}
   Dk=\omega\,d\omega\,d{\bf n}\,,
\end{equation}
where $d{\bf n}$ is the restriction of the Lebesgue measure in the
three dimensional sphere. Using the fact that the function
$\phi(k_i,\rho)$ is homogenous of the degree $-1+i\rho$, we can
derive the following expression:

\begin{equation}\label{3.84}
f(x_i)= \frac{1}{16\pi^3}\int_{-\infty}^{\infty}d\rho\, \rho^2\int
\omega\,d\omega\, d{\bf
n}\,\phi(N_i,\rho)\,\omega^{-1+i\rho}\,\delta(x_i N^i\omega-1)\,.
\end{equation}
If we integrate (\ref{3.84}) with respect to $\omega$, one obtains

\begin{equation}\label{3.85}
  f(x_i)= \frac{1}{16\pi^3}\int_{-\infty}^{\infty}d\rho\,
  \rho^2\int d{\bf n}\, \phi(N_i,\rho)(x_i N^i)^{-1-i\rho}\,.
\end{equation}

Relation (\ref{3.85}) gives the inverse Fourier transform of
$f(x_i)$ in terms of the eigenfunctions $(x_i N^i)^{-1-i\rho}$ of
the Hamiltonian, i.e., what we have called the hyperbolic plane
waves or Shapiro waves \cite{S,VS}. For real $\rho$, the set of
functions $(x_i N^i)^{-1-i\rho}$ realize a unitary representation
for the principal series of $SO(3,1)$.

\section{Concluding remarks}

We have investigated the possibility of constructing Spectrum
Generating Algebras (SGA) for quantum systems showing a purely
continuous spectrum. In fact, we have obtained a SGA for the free
particle in the three dimensional two sheeted hyperboloid ${\cal
H}^3$. We have done this in two steps. First of all, we have
obtained a representation of the Lie algebra $\frak{so}(4,2)$, by
functions of coordinates and momenta, suitable for the description
of a classical particle on an one or two sheeted hyperboloid. We
have obtained the Dirac-Poisson brackets for the generators of the
algebra. In the second step, we have obtained another representation
of $\frak{so}(4,2)$ in which functions are replaced by operators and
Dirac brackets by commutators. Following a usual procedure, we
construct ladder operators as Hermitian members of the algebra
$\frak{so}(4,2)$.

We have found the solutions of the Schr\"odinger equation in ${\cal
H}^3$ equivalent to the plane waves in the space ${\mathbb R}^3$. As
solutions of a time independent Schr\"odinger equation, these plane
waves are eigenvalues of the Hamiltonian, so that they can be
labeled by their energies. Instead, we prefer to label them by the
eigenvalues of a related operator $h$ as given in (\ref{3.40}),
which is one of the generators of the algebra. We denote by $\rho$
to the eigenvalues of $h$.

We observe that the ladder operators shift the variable $\rho$ in
these solutions by a complex number. This may happen because our
generalized plane waves are out of the Hilbert space. In order to
avoid this inconvenience, we have introduce some operators which are
functions of suitable linear combinations of the ladder operators.
These operators are not self adjoint but produce real shifts on the
label $\rho$ of the generalized plane waves and can be used as a new
form of ladder operators for the continuous spectrum.

Finally, we have discussed a generalized Fourier transform between
functions on the three dimensional hyperboloid and functions over a
three dimensional cone. This is intimately related to the
transformation defined by Graev and Gelfand in \cite{GGV}. A
Plancherel type theorem is valid in this context. We have given an
eigenfunction expansion of functions over the hyperboloid in terms
of the generalized plane waves on the hyperboloid.

\section*{Acknowledgements}

 Partial financial support is indebt to the  Ministry of Science
(Project MTM2009-10751 and FIS2009-09002), and to the Russian
Science Foundation (Grants 10-01-00300).


\begin{thebibliography}{99}

\bibitem{BB} A.O. Barut, A. Bohm, {\it Dynamical groups and mass formula} Phys. Rev, {\bf 139}, B1107
(1965).

\bibitem{D} Y. Dothan, M. Gell-Mann, Y. Ne'eman,
{\it Series of hadron energy levels as representations of
non-compact groups} Phys. Lett, {\bf 17}, 148 (1965).

\bibitem{DI} P.A.M. Dirac, {\it  Lectures on Quantum Mechanics},  (Belfer Graduate School of
Science Monographs Series Number 2, 1964).



\bibitem{guru} E.C.G. Sudarshan, N. Mukunda, {\it Classical Dynamics:
A Modern Perspective} (Wiley, New York, Toronto, 1974).



\bibitem{P1} M. Gadella, J. Negro, L.M. Nieto, G.P. Pronko and M. Santander,
{\it Spectrum Generating Algebra for the free motion in $S^3$},
Journal of Mathematical Physics, {\bf 52}, 063509 (2011).

\bibitem{GGV} I.M. Gelfand, M.I. Graev, N. Ya. Vilenkin, {\it
Generalized Functions; Integral Geometry and Representation Theory},
(Academic, New York and London 1966).





\bibitem{O} Y. Ohnuki, {\it Unitary representations of the
Poincar\'e group and relativistic wave equations} (World Scietific,
Singapore, 1988).

\bibitem{GG} M. Gadella, F. G\'omez, {\it A unified mathematical formalism for the Dirac
formulation of quantum mechanics}, Foundations of Physics, {\bf 32},
815-869 (2002); M. Gadella, F. G\'omez-Cubillo, {\it Eigenfunction
Expansions and Transformation Theory}, Acta Applicandae
Mathematicae, {\bf 109}, 721-742 (2010).

\bibitem{K} A.A. Kirilov, {\it Elements of the Theory of
Representations} (Springer, Berlin, Heidelberg, New York, 1976).

\bibitem{AJS} W.O. Amrein, J.M. Jauch, K.B. Sinha, {\it Scattering
Theory in Quantum Mechanics} (Benjamin, Reading, 1977).

\bibitem{guru1} J.G. Kuriyan, N. Mukunda, E.C.G. Sudarshan, {\it Master
analytic representation-reduction of $O(2,1)$ basis}, Journal of
Mathematical Physics, {\bf 9}, 2100 (1968); N. Mukunda, {\it Unitary
representations of group $O(2,1)$ in an $O(1,1)$ basis }, Journal of
Mathematical Physics, {\bf 8}, 2210 (1967).

\bibitem{GR} I.S. Gradshteyn, I.M. Ryzhik, {\it Table of integrals,
series and products} (Academic, New York 1965).

\bibitem{S} I. S. Shapiro, {\it  Development of a wave function into
a series by irreducible representations of the Lorenz group},
Doklady Akademii Nauk, {\bf 106} 647-649 (1956).

\bibitem{VS} N. Ya. Vilenkin, Ya. A. Smorodinskii, {\it  Invariant expansions
of relativistic amplitudes} Soviet Physics JETP, {\bf 19},
1793 (1964).




\end{thebibliography}
\end{document}